\documentclass[aps,prl,preprint,showpacs,amssymb,groupedaddress]{revtex4}
\usepackage{graphicx}

\begin{document}


\title{Velocity correlations in dense granular flows observed with
  internal imaging}

\author{Ashish V. Orpe and Arshad Kudrolli} \affiliation{Department of
  Physics, Clark University, Worcester, Massachusetts 01610}

\date{\today}

\begin{abstract}
  We show that the velocity correlations in uniform dense granular
  flows inside a silo are similar to the hydrodynamic
  response of an elastic hard-sphere liquid.  The measurements are
  made using a fluorescent refractive index matched interstitial fluid
  in a regime where the flow is dominated by grains in enduring
  contact and fluctuations scale with the distance traveled, independent
  of flow rate. The velocity autocorrelation function of the grains in
  the bulk shows a negative correlation at short time and slow
  oscillatory decay to zero similar to simple liquids. Weak spatial
  velocity correlations are observed over several grain diameters. The
  mean square displacements show an inflection point indicative of
  caging dynamics. The observed correlations are qualitatively
  different at the boundaries.
\end{abstract}



\maketitle

The structure and dynamics of dense granular flows is a problem of
fundamental interest. Investigating the correlations in the
fluctuations of the grain motion can yield insight into the caging and
diffusion of particles. 
And as measured by the velocity auto-correlation functions, they can
also illustrate memory in the motion of the constituents. Having a
measure of the correlations is important in developing a hydrodynamic
description for granular materials and interpretation of variables used
to characterize their properties.

Considerable relevant work exists in the context of simple fluids
modeled as elastic hard spheres~\cite{hansen91}. Computer simulations
have shown that the velocity autocorrelation function of dense elastic
particles in equilibrium exhibit a negative correlation at short times
due to backscatter~\cite{rahman64}. The resulting coupling of the
tagged particle to the hydrodynamic modes are implicated in the
observation of power law decays of velocity~\cite{alder70}, which is
in contrast with the anticipated exponential decay based on Markovian
interactions of particles with neighbors. These effects were later
described by hydrodynamic and mode-coupling
theories~\cite{zwanzig70,pomeau75}, and are said to be in excellent
agreement with direct observations in colloidal
systems~\cite{megen02,weeks02}. Thus correlations are observed even in simple dense liquids where particles interact elastically.

In granular systems, because inelasticity and friction is important whenever
grains come in contact, relative motion is suppressed unless energy is
supplied externally. Thus it is unclear if correlations present in
granular systems are similar to simple liquids~\cite{dufty03}.
Recently, fluctuations and correlations have been reported in dense
granular systems, but the observations without exception have been
made next to the sidewalls or at the free
surface~\cite{menon97,mueth03,choi04,pouliquen04,moka05}.  The source for
correlations are complicated by the direct influence of the boundary
on caging and diffusion, and the presence of shear in the location
where the observations are made.

Here, we examine simplified uniform dense granular flows which occur
inside a silo away from the side walls to compare and contrast their
fluctuation properties with simple liquids in equilibrium. We use a
fluorescent refractive index matched liquid
technique~\cite{tsai03,siavoshi06} to measure the correlations in
motion of the grains in the bulk, and further compare it with those
near the sidewalls. The fluctuation properties of the particles are
observed to be independent of the flow rate, and therefore the
interstitial fluid has no significant effect on the grain fluctuations
in our experiments. We find that the correlations in the granular
fluctuations as measured by the mean square displacements, and the
velocity auto-correlation function are remarkably similar to that
observed in elastic hard sphere liquids. Thus our measurements support
hydrodynamic approaches for granular systems using approximations made
for many-body correlations as for simple liquids.

\begin{figure}
  \includegraphics[width=1.0\linewidth]{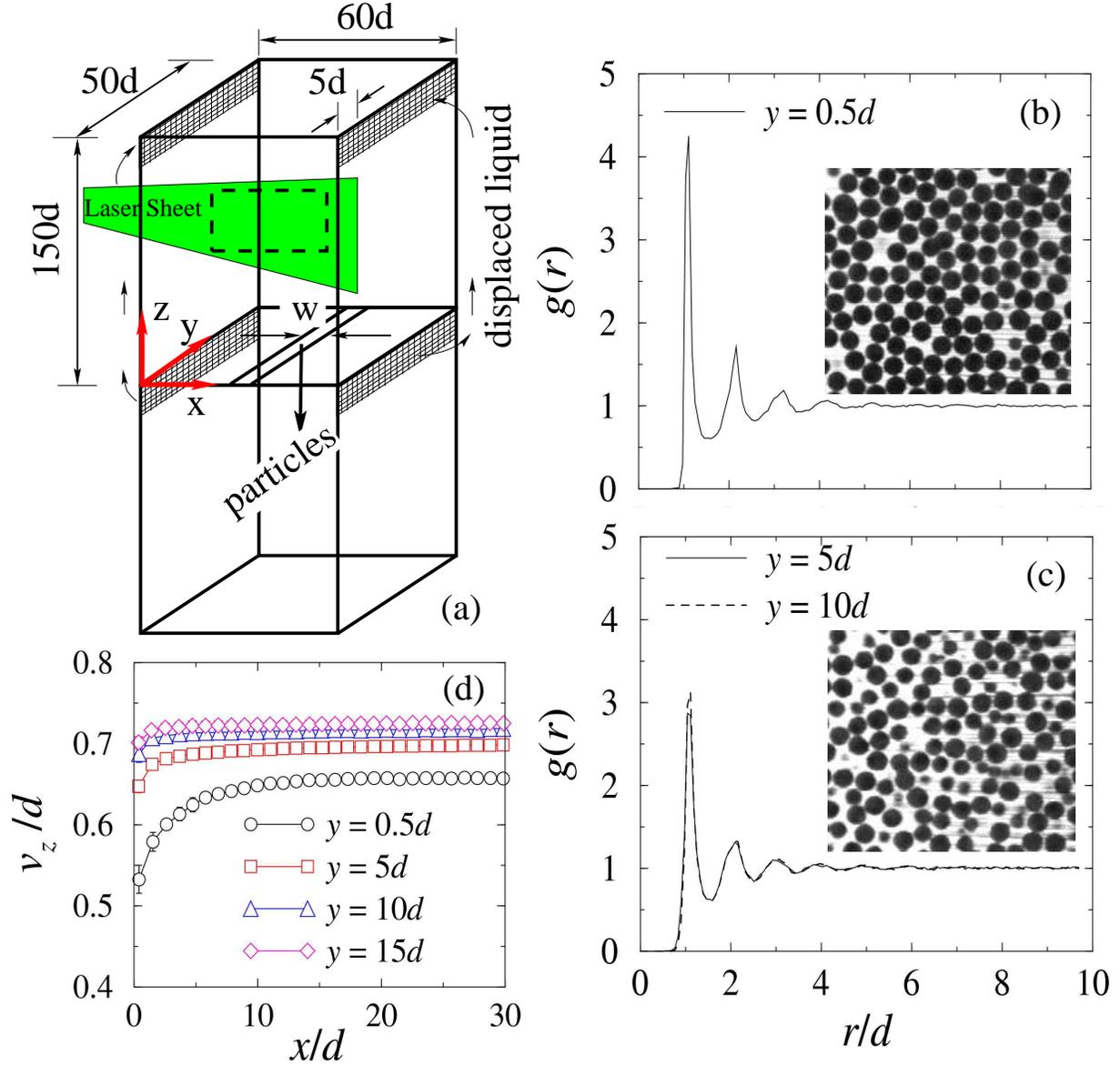}
\caption{
  (a) The schematic diagram of the silo apparatus. A laser sheet
  illuminates a cross section of particles in a plane at a distance
  $y$ from the front surface.  (b) The two-point spatial correlation
  function $g(r)$ of the grains located next to the front wall of the
  silo ($y=0.5d$).  Inset: A typical image of the grains.  (c) $g(r)$
  in a plane $y = 5d$ and $y = 10d$.  Inset: A typical image for $y =
  5d$.  (d) The vertical component of the velocity $v_{z}$ as a
  function of distance $x$ along the horizontal axis at $y = 0.5d,
  5d$, $10d$ and $15d$ ($w = 4.0d$). Error bars show the deviations
  over 10 runs.}
\label{fig1} 
\end{figure} 

The experimental apparatus consists of a glass silo chamber with
dimensions shown in Fig.~\ref{fig1}(a). Glass beads with diameter $d =
1 \pm 0.1$\,mm drain from the silo into a bottom collecting chamber through an exit slot. The flow rate is set by varying the width $w$ of the slot. A
minimum width ($w =3.25d$) was needed to observe steady flow, and the
data reported here was obtained for $w = 3.5d, 4d, 5d$ and $6d$.
To measure the flow away from the sidewalls, the
entire system is immersed in an interstitial fluid~\cite{liquids}
with the same refractive index ($\approx$ 1.52) as the glass beads.  The
fluid displaced in the bottom chamber is channeled through a mesh via
side chambers (not shown) into the silo at the top. This arrangement
was found to effectively reduce counter flow of the interstitial
fluid through the exit slot.

The grains are visualized by adding a fluorescent dye to the
fluid~\cite{siavoshi06}.  As illustrated in Fig.~\ref{fig1}(a), a
plane inside the silo, which is less than 0.1d thick, is illuminated using a $50$-mW laser and a cylindrical lens, and imaged through the front wall with a digital
camera.  Typical images for two different planes, wherein the
particles appear dark against a bright background, are shown in insets
to Fig.~\ref{fig1}(b),(c). The apparent size of the particles depends on
the distance of the bead center from the laser illumination plane. A centroid algorithm is used to find sub-pixel resolution particle position with centers within $\pm$ 0.3d of the laser sheet. Imaging at $60$ frames per second is sufficient to track the particles over long time periods and obtain mean velocities to within $1\%$.

The granular flow in the silo can be divided into two distinct
regions: a convergent accelerating flow close to the exit slot and a
steady plug-like flow with small shear at the walls in the region
above $z = 70d$. The competition between grains exiting from the orifice results in packing density fluctuations which diffuse up through the silo as grains fall down and has been discussed previously~\cite{mullins}. As we are primarily interested in the
fluctuation properties of the flow, we focus on the relatively simpler
uniform flow region at the top.

We first examine the packing structure of the particles at various
locations in the silo. The grain packing as seen in typical images
corresponding to $y = 0.5d$, and $y = 5d$ in Fig.~\ref{fig1}(b),(c)
are qualitatively different. In order to quantitatively measure the
difference, we calculated the two-point spatial correlation function
$g(r)$ as a function of particle separation distance $r$ in a $y$
plane for $y=0.5d, 5d$, and $10d$ [see Fig.~\ref{fig1}(b),(c)]. While
$g(r)$ corresponding to $y = 0.5d$ show stronger peaks, $g(r)$ for $y
= 5d$ and $y =10d$ are indistinguishable and the packing structure does
not change significantly any further distance from the boundaries. The
packing fraction of the beads inside the silo is measured to be
$0.5965 \pm 0.0123$ and is observed to be independent of flow rate.
Furthermore, $g(r)$ in the flowing regions is similar to that measured when the grains are at rest.

The vertical component of the velocity $v_{z}$ as a function of
horizontal distance is plotted in Fig.~\ref{fig1}(d) for grains at
various distances from the front wall. Because the flow is symmetric
about $x=30d$, we show only the left half. The velocity increases with
distance away from the sidewalls over a distance of about $10d$,
beyond which it remains more or less constant.

\begin{figure}
  \includegraphics[width=1.0\linewidth]{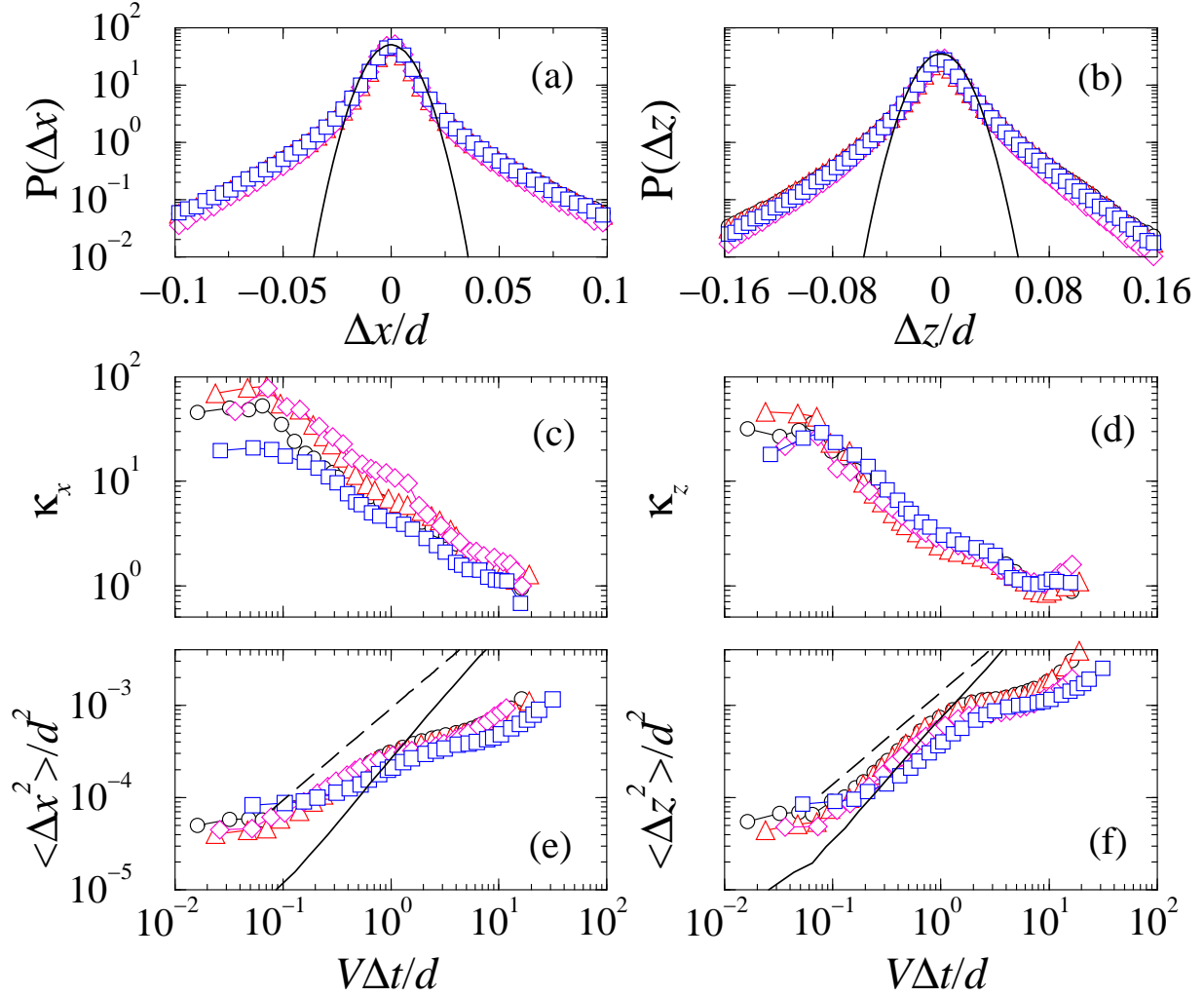}
\caption{(a, b) Probability distribution function (PDF) of particle
  displacements, $\Delta x$ and $\Delta z$, measured over a time
  interval $\Delta t = 1d/V$ in the plane $y = 10d$. {\large$\circ$}: $V =
  0.5d/s$, {\scriptsize$\triangle$}: $V = 0.7d/s$, {\large$\diamond$}:
  $V = 1.1d/s$, {\tiny$\square$}: $V = 1.6d/s$, Gaussian distribution
  (solid line).  (c,d) The kurtosis of the $\Delta x$ and $\Delta z$
  PDFs versus the mean distance traveled. (e,f) The horizontal and
  vertical mean square displacement versus mean distance traveled by
  the grains. Dashed line of slope $1$ corresponds to a purely
  diffusive flow. Solid line correspond to $y=0.5d$ and $V = 0.7d/s$.
}
\label{fig2}
\end{figure}

In the subsequent analysis, we focus in the central nearly shear free
region, of size $20d$ across and $20d$ in the flow direction, where
the velocity magnitude varies within $2\%$. The fluctuation properties
are then computed in this region for each plane $y$ from the
trajectories of individual particles. The horizontal particle displacement $\Delta x = x(t+\Delta t) - x(t)$ and the vertical particle displacement $\Delta z  = z(t+\Delta t) -z(t) - V \Delta t$ in the reference frame of
the flow are determined over a time interval $\Delta t$, and 
time averaged uniform mean flow $V$ at the silo center ($x=30d$). 
The normalized probability density functions (PDF) for $\Delta x $ and $\Delta z$ in the plane $y=10d$ for  $ \Delta t = d / V$ are shown in Fig.~\ref{fig2}(a) and
(b), respectively, for different flow rates. The PDFs are nearly same for the various
flow rates, and are non-Gaussian similar to observations in dry
granular silo flows near the side walls~\cite{choi04,moka05}. The
ratio of mean square deviations of the PDFs in the vertical to the
horizontal directions is 1.33, and is similar to that in dry granular
silo systems~\cite{choi04}. The anisotropy arises because energy is fed into the system by grains falling down under gravity and lost due to dissipative collisions with neighbors in all directions. Non-Gaussian PDFs have been observed
in colloidal glasses~\cite{weeks02} as well and are attributed to cage
breaking and depends on the time interval $\Delta t$ used to calculate
the displacements.

The deviation of the PDFs from a Gaussian distribution is determined
using the normalized kurtosis, defined as $\kappa_{\alpha} =
\;<\!\Delta \alpha^{4}>\!\!/3\!\!<\!\Delta \alpha^{2}\!>^{2}
- 1$, where $\alpha = x, z$. The kurtosis for $\Delta x$ and $\Delta
z$ as a function of normalized $\Delta t$ is shown in
Fig.~\ref{fig2}(c) and (d), respectively.  In all cases, kurtosis
decreases inversely with time to zero indicating an approach to diffusive
motion as grains undergo several rearrangements.

To characterize the fluctuations further, we plot the mean square
horizontal $<\!\!\Delta x^{2}\!\!>$ and vertical $<\!\!\Delta
z^{2}\!\!>$ displacements in the flow frame of reference versus distance traveled with  the flow (and
thus time) in Fig.~\ref{fig2}(e,f). Above the noise floor, we find
that $<\!\!\Delta z^{2}\!\!>$ increase approximately linearly, shows
an inflection point, and then curves up to approach a linear increase
corresponding to diffusive motion. $<\!\!\Delta x^{2}\!\!>$ is
somewhat lower in comparison with $<\!\!\Delta z^{2}\!\!>$ just as for
the PDFs. The inflection point and linear to linear behavior seen here
is qualitatively similar to observations in elastic hard sphere
suspensions for densities near the glass
transition~\cite{megen02,weeks02}. There the inflection point is
associated with a slow coherent cage breaking motion that takes
longer time to develop and diverges as the volume fraction is
increased. The mean square displacements at the boundaries [solid lines in
Fig.~\ref{fig2}(e),(f)] show a very different behavior 
(slope $\approx$ 1.4). But it is noteworthy that it is in agreement with the experimental results measured for dry granular flows at the boundaries~\cite{choi04,moka05}.
 
\begin{figure}
  \includegraphics[width=1.0\linewidth]{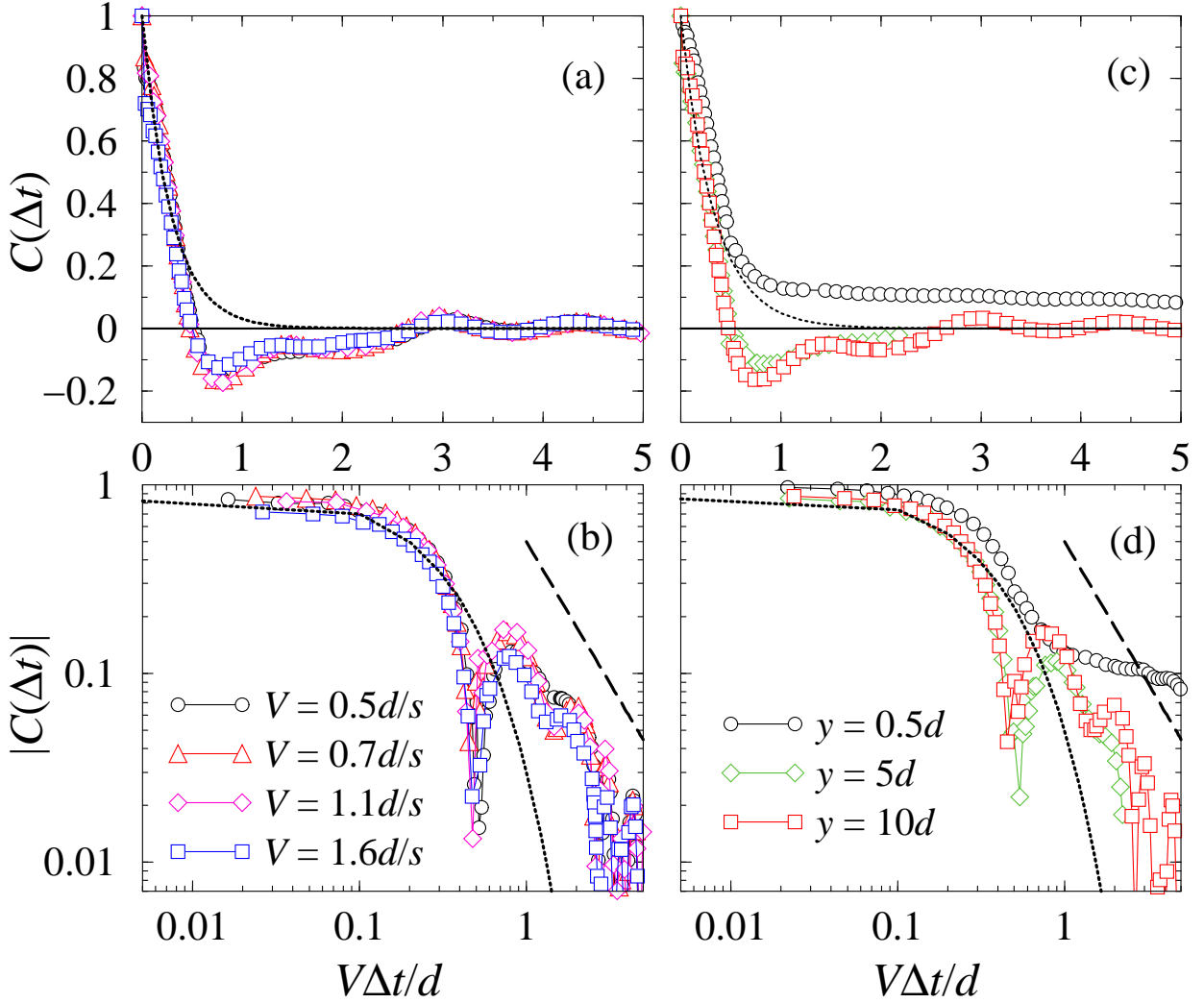}
\caption{The velocity auto-correlation function $C(\Delta t)$.  
(a,b) Effect of flow rate. Data for the plane $y = 10d$.
  and (c,d) Effect of end walls. Data for mean flow velocity $V =
  0.7 d/s$.  Velocities calculated using a time interval 
  $\Delta t = 0.5d/V$. Dotted line represents an exponential
  decay while dashed line represents long time power law decay
  ($\Delta t^{-3/2}$).} 
\label{fig3} 
\end{figure}

To gain further information on the cooperative nature of the
fluctuations, we next turn to velocity correlations and examine the
velocity auto-correlation function (VAF). For grains in a given $y$
plane, the instantaneous fluctuating velocity components ($c_{x}$, $c_{z}$) at any time instant are obtained by subtracting the mean flow velocity at that instant from the individual velocities of the grain. The VAF is, then, calculated from the measured velocities using the definition $C(\Delta t) = <\!\!c_{x}(t)c_{x}(t+\Delta t) + c_{z}(t)c_{z}(t+\Delta t)\!\!>$, where the angled brackets represents the averaging over all times $t$ of
flow.  The VAFs for $y = 10d$ is shown using linear scale in
Fig.~\ref{fig3}(a) and double-log scale in Fig.~\ref{fig3}(b). The
profiles are distinctly different from an exponential decay (dotted
line) displayed by a Brownian particle obeying the Langevin equation.
In particular, it may be noted that VAF becomes negative over a time
scale comparable to the time taken for the grains to flow about a
grain diameter, irrespective of the flow rate. Then, the VAF oscillates and 
decays to zero over long times. The magnitude of the negative
correlation and the overall shape of the decay is very similar to that
observed first by Rahman~\cite{rahman64} in simulations of liquid
Argon, and more recently in the molecular dynamics simulations of
meta-stable liquids at high volume fractions by Williams, et al~\cite{williams06}.

The negative correlations and slow decay in dense liquids are said to
arise due to back-scatter of the tagged particles and reversal of
velocity into a comparatively narrow range of angles, and development
of a back flow~\cite{hansen91}. 
Because many particles are involved in this process, a hydrodynamic
model was developed to describe the
correlations~\cite{zwanzig70,alder70}. The model gives a 3/2 scaling
for the decay of the correlations due to the diffusion of momentum of
the tagged particle. This scaling is denoted by the dashed line in
Fig.~\ref{fig3}(b,d). Although the data is consistent with 3/2 scaling, our observation time window is rather limited.

The data corresponding to grains next to the front wall and at various
depths $y$ is shown in Fig.~\ref{fig3}(c,d). The behavior is
qualitatively different at the walls and negative VAFs are not
observed. It is possible that differences can arise for the following reasons. The particles in the bulk can move in all three directions, so the fluctuations diffuse faster than those at the walls which limit grain fluctuation in that direction. Further, as can be seen from Fig.~1(b,c), the grains near the wall appear more structured which may also lead to differences. The boundaries also induce shear and Kumaran~\cite{kumaran06} has recently
argued that VAFs can decay faster in sheared granular flow than for fluids in equilibrium. However, these effects appear to be not so strong when comparing the results at the boundaries.

\begin{figure}
  \includegraphics[width=0.7\linewidth]{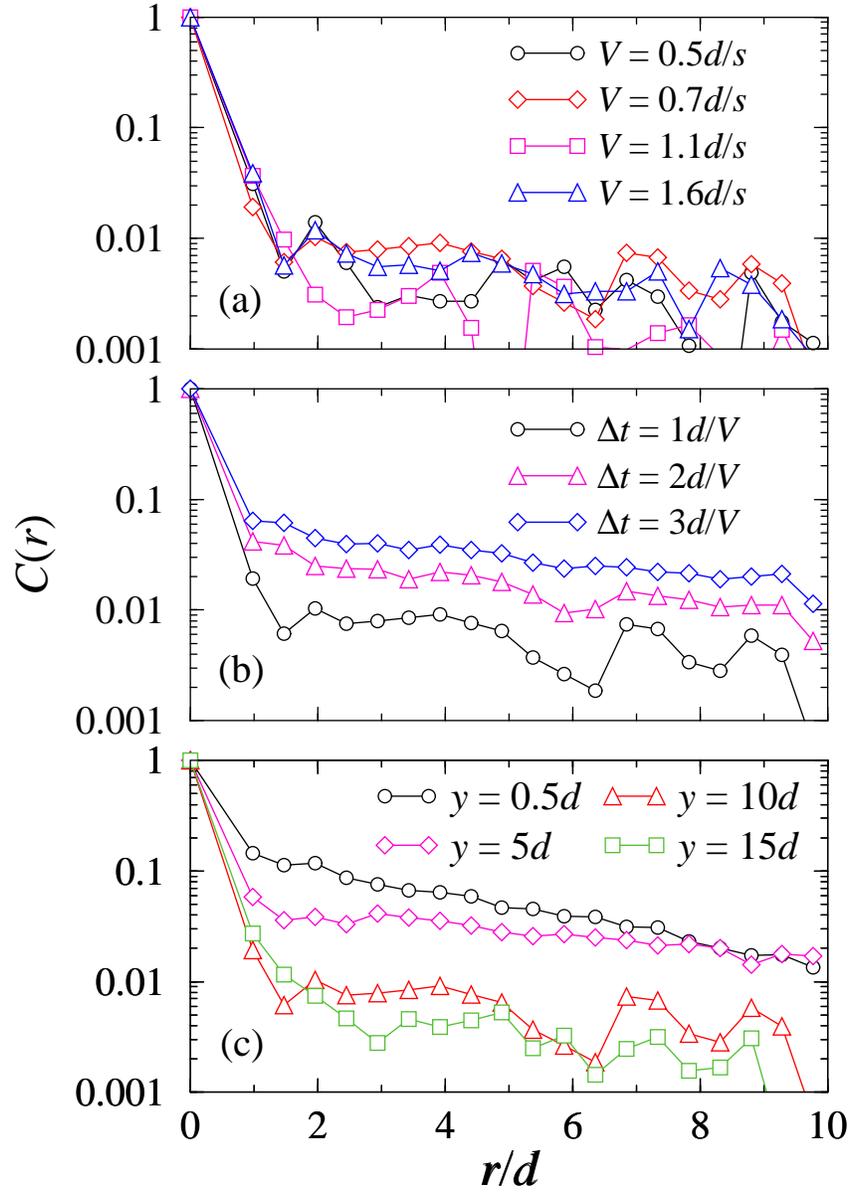}
\caption{Spatial velocity correlation $C(r)$ versus the
  interparticle distance $r/d$. (a) Effect of flow rate. ($\Delta t
  =d/V$) (b) Effect of the averaging time interval. ($y = 10d$ and $V
  = 0.7 d/s$) (c) Effect of the side walls.  ($\Delta t =d/V$ and $V
  = 0.7 d/s$) .}
\label{fig4} 
\end{figure}

The implied velocity fields can be tested statistically using the
spatial velocity correlation function defined as $C(r) = <\!\!c_{x}({\bf r_0})c_{x}({\bf r_0+r}) + c_{z}({\bf r_0})c_{z}({\bf r_0+r})\!\!>$.  The angled brackets represents the averaging over all particle centered at ${\bf r_0}$ and over all the measurement time window for particles separated by a distance $r$.
The measured $C(r)$ for various flow rates is shown in
Fig.~\ref{fig4}(a) where the velocity has been calculated over a time
interval $\Delta t = d/v_z$ corresponding to the minima in $C(\Delta t)$.
(Separate correlations for the $x$ and $z$ components were found
to be similar, and therefore the averaging over $r$ is reported here
to consolidate the graphs.) Only weak spatial velocity correlations
close to the noise floor may be noted over this time scale. $C(r)$
increases modestly if the velocity is calculated over a longer time
interval corresponding to the slow decay observed in $C(\Delta t)$ [see
Fig.~\ref{fig4}(b)]. Therefore while a signal of cooperative flow is
observed, it appears to be rather weak perhaps because the cooperative motion is spread
over a large number of particles.

To compare the spatial correlations observed in the bulk with those
at the boundaries, $C(r)$ is plotted for various planes from the front
wall 
(see Fig.~\ref{fig4}(c)). It can be noted that significantly
greater spatial correlations are observed near the boundaries. The
strength of correlations is in fact similar to that observed in dry
granular flows where observations have been made only next to the side
walls and where shear is present~\cite{moka05,mueth03}. Thus the
correlations as measured by mean square deviations, $C(\Delta t)$, and $C(r)$ are consistently
different in the shear free bulk flow regions and at the boundaries.
Further work is required to distinguish between the contribution due
to the screening and ordering induced by the boundaries, and the
contribution due to shear.

Finally, we discuss the effect of the interstitial fluid on the
reported fluctuation properties of the grains. In principle, viscous
forces due to the interstitial fluid exist between particles which can
affect their fluctuations. These forces are directly proportional to
the relative velocity and thus rate dependent. However, as the PDFs, 
$<\!\!\Delta x^{2}\!\!>$, $<\!\!\Delta z^{2}\!\!>$, and $C(\Delta t)$ 
coincide when plotted over distance traveled in the flow,
irrespective of flow rate, we conclude that any rate dependent forces
introduced by the interstitial fluid has insignificant effect on
the fluctuation properties discussed here. This argument, along with our observations that the fluctuation properties near the side walls are consistent with that of dry granular systems reported previously, indicates that our results are applicable to dry
granular flows in the bulk as well.

In conclusion, the correlations in fluctuations observed for grains
undergoing uniform flow are remarkably similar to that exhibited by an
elastic hard-sphere liquid. In hydrodynamic models of elastic
hard-spheres, the Enskog equation is used to calculate transport
coefficients, which includes the finite size of the particles but
ignores the velocity correlations built up by successive collisions~\cite{hansen91}.
Our observations seem to indicate that the dissipative nature of the
grain-grain interaction do not significantly change the corresponding correlations in granular systems, and suggests that a similar approach may be fruitful for dense granular flows.

\acknowledgments{
We thank J. Norton for his help with the apparatus. This work was
supported by the National Science Foundation under grant number CTS-0334587, and the
donors of the Petroleum Research Fund.}



\begin{thebibliography}{0}
 
\bibitem{hansen91} J. P. Hansen and I. R. McDonald, {\it Theory of
    Simple Liquids} (Academic Press, New York, 1991).
  
\bibitem{rahman64} A. Rahman, Phys. Rev. {\bf 136}, A405 (1964).
  
\bibitem{alder70} B. J. Alder and T. E. Wainwright, Phys. Rev. A {\bf
    1}, 18 (1970).
  
\bibitem{zwanzig70} R. Zwanzig and M. Bixon, Phys. Rev. A {\bf 2},
  2005 (1970).
  
\bibitem{pomeau75} Y. Pomeau and P. Resibois, Phys. Rep. {\bf 19}, 63
  (1975).

  
\bibitem{weeks02} E. R. Weeks and D. A. Weitz, Phys. Rev. Lett. {\bf
    89}, 095704 (2002).
  
\bibitem{megen02} W. van Megen, J. Phys.: Condens. Matter {\bf 14},
  7699 (2002).
  
\bibitem{dufty03} J. W. Dufty and J. J. Brey, Phys. Rev. E {\bf 68},
  030302(R) (2003).
  
\bibitem{menon97} N. Menon and D. J. Durian, Science {\bf 275}, 1920
  (1997).
  
\bibitem{mueth03} D. M. Mueth, Phys. Rev. E {\bf 67}, 011304 (2003).
  
\bibitem{choi04} J. Choi, A. Kudrolli, R. R. Rosales, and M. Z.
  Bazant, Phys. Rev. Lett. {\bf 92}, 174301 (2004).
  
\bibitem{moka05} S. Moka and P. R. Nott, Phys. Rev. Lett. {\bf 95},
  068003 (2005).
  
\bibitem{pouliquen04} O. Pouliquen, Phys. Rev. Lett. {\bf 93}, 248001
  (2004).
  
\bibitem{siavoshi06} S. Siavoshi, A. V. Orpe, and A. Kudrolli, Phys.
  Rev.  E {\bf 73}, 010301(R) (2006).
  
\bibitem{tsai03} J.-C. Tsai, G. A. Voth, and J. P. Gollub, Phys. Rev.
  Lett. {\bf 91}, 064301 (2003).
  
\bibitem{liquids} The liquids were obtained from Cargille
  Laboratories, and have a density of $1100$ kg m$^{-3}$  and viscosity
  of $2 \times 10^{-5}$ m$^{2}$ s$^{-1}$.
  
\bibitem{mullins} J. Mullins, J. Appl. Phys. {\bf 43}, 665 (1972); M. Z. Bazant, Mech. Mater. {\bf 38}, 717 (2006).

\bibitem{williams06} S. R. Williams, G. Bryant, I. K. Snook, and W.
  van Megen, Phys. Rev. Lett. {\bf 96}, 087801 (2006).
  
\bibitem{kumaran06} V. Kumaran, Phys. Rev. Lett. {\bf 96}, 258002
  (2006).

\end{thebibliography}
\end{document}